\documentclass{svjour3}                     
\smartqed  
\usepackage{graphicx}
\usepackage[dvips]{color}
\usepackage{graphicx}
\usepackage{dcolumn}
\usepackage{amsmath}
\usepackage{amssymb}
\usepackage{amsfonts}
\def\beq{\begin{equation}}
\def\eeq{\end{equation}}
\def\beqa{\begin{eqnarray}}
\def\eeqa{\end{eqnarray}}

\begin{document}

\title{Perturbative treatment of three-nucleon force contact terms in
  three-nucleon Faddeev equations}


\titlerunning{Perturbative treatment of three-nucleon force contact ....}      

\author{H.~Wita{\l}a, J.~Golak, R.~Skibi\'nski, K.~Topolnicki}

\authorrunning{H. Wita{\l}a, J. Golak, R. Skibi\'nski, K. Topolnicki}

\institute{H. Wita{\l}a  \at
              M. Smoluchowski Institute of Physics, Jagiellonian
 University, PL-30348 Krak\'ow, Poland\\
              \email{henryk.witala@uj.edu.pl}
\and
J. Golak  \at
              M. Smoluchowski Institute of Physics, Jagiellonian
              University, PL-30348 Krak\'ow, Poland\\
\and
R. Skibi\'nski  \at
              M. Smoluchowski Institute of Physics, Jagiellonian
              University, PL-30348 Krak\'ow, Poland\\
\and
K. Topolnicki  \at
              M. Smoluchowski Institute of Physics, Jagiellonian
              University, PL-30348 Krak\'ow, Poland
}

\date{Received: date / Accepted: date}

\maketitle

\begin{abstract}
  We present a perturbative approach to solving the three-nucleon continuum
  Faddeev
  equation. This approach is particularly
  well suited to dealing with variable strengths of contact terms in a chiral
  three-nucleon force.  
 We use examples of observables in
  the elastic nucleon-deuteron scattering as well as in the deuteron 
  breakup reaction to demonstrate high precision of the proposed procedure 
 and its capability to reproduce exact results.
A significant reduction of computer time achieved 
by the perturbative approach in comparison to exact treatment makes this 
approach 
  valuable for fine-tuning of the three-nucleon Hamiltonian parameters. 
\end{abstract}


\vskip 0.5 true cm

\noindent{Special Issue:
“Celebrating 30 years of Steven Weinberg's papers on Nuclear Forces from Chiral
  Lagrangians"}

\section{Introduction}
\label{intro}

 The nuclear force problem is the basic one for understanding nuclear
 phenomena. Since the birth of nuclear physics it has been at the centre of
 experimental and theoretical studies. Extensive efforts based 
 on purely phenomenological approaches or incorporating
 the meson-exchange picture have led to
 numerous nucleon-nucleon (NN) potentials, able to describe with high
 precision  a vast
 amount of available data \cite{mach_anp}. In spite of the enormous progress
 in understanding properties of the two-nucleon interaction, applications of
 these ideas to the many-nucleon forces encountered consistency
 problems and called for a more systematic framework.
 The advent of QCD gave a new impetus to the derivation of nuclear forces
 and a major
 breakthrough occurred with the emergence of the effective field theory (EFT)
 concept
 and publishing of S. Weinberg's seminal paper \cite{weinberg}. It paved
 the way
 for developing accurate and precise nuclear forces within the EFT
 framework \cite{vankolck,epel_nn_n3lo,epel6a,Epelbaum:2008ga,machl6b}.

 The progress in constructing nuclear forces within EFT approach
 is presently documented by availability
 of  numerous high precision NN potentials which provide satisfactory
 description of NN data in a wide energy range.  
 Recently a new generation of chiral NN potentials
 was introduced and developed up to the fifth order (N$^4$LO) of
 chiral expansion
 by the
Bochum-Bonn \cite{epel1,epel2} and Idaho-Salamanca \cite{entem2017} groups.
While in the Idaho-Salamanca force the nonlocal momentum space regularization
is applied directly in momentum space by introducing a cutoff 
parameter $\Lambda$, 
the one-pion  and two-pion exchange contributions 
in the Bochum-Bonn potential
are regularized in coordinate space using a cutoff parameter $R$
and then transformed to momentum space.
 For the contact interactions the Bochum-Bonn NN force employs 
 a simple Gaussian nonlocal 
 momentum-space regulator with the cutoff $\Lambda = 2R^{-1}$.
 Potentials of both groups are available for a number of regularization
 parameters. They  provide  a very
good description of the NN data set (Idaho-Salamanca) or the phase shifts and
mixing angles of the Nijmegen partial wave analysis \cite{nijmpwa}
(Bochum-Bonn), 
used to fix the low-energy constants accompanying the NN contact
interactions. 
The latest and most precise EFT-based NN interaction is the so called semilocal
momentum-space (SMS)
regularized chiral potential of the Bochum-Bonn group \cite{preinert},
 developed up to N$^4$LO and even including some terms from the next order
 of chiral expansion (N$^4$LO$^+$).
 In this potential a new momentum-space regularization scheme
 has been employed 
 for the long-range contributions 
 and a nonlocal
 Gaussian regulator has been applied to the minimal set of independent
 contact interactions. 
 This new approach can be straightforwardly utilised to regularize
 also three-nucleon (3N) forces. 
 That new family of semilocal chiral potentials provides an outstanding
 description of the NN data and is available up to N$^4$LO$^+$
 for five values of the cutoff $\Lambda$.

 Applications of EFT approach in the form of chiral
 perturbation theory (ChPT) have resulted not only in the theoretically
 well grounded
 NN potentials but also for the first time have given a
 possibility to apply in practical calculations NN forces augmented with 
 consistent 3N interactions, both derived within the same  formalism.
 Understanding of nuclear spectra and reactions based on these 
  consistent chiral two- and many-body forces has become a hot topic
  of present day few-body studies and is also the main aim of the
  Low Energy Nuclear Physics International Collaboration
  (LENPIC) \cite{epel2019}.

  The first nonvanishing contributions to the 3N force (3NF) appear
  at next-to-next-to-leading order of chiral expansion (N$^2$LO)
 \cite{vankolck,epel2002} and comprise
in addition to the $2\pi$-exchange term two contact contributions with strength
parameters $c_D$ and $c_E$ \cite{epel_tower}. 
 The difficult task to derive the chiral
3NF at next-to-next-to-next-to-leading order (N$^3$LO) 
has been performed in \cite{3nf_n3lo_long,3nf_n3lo_short}.
 At that order five different topologies contribute
 to 3NF. Three of them are of long-range character \cite{3nf_n3lo_long}
 and are given by two-pion ($2\pi$) exchange graphs, by
 two-pion-one-pion ($2\pi-1\pi$) exchange graphs, and
by the so-called ring diagrams. They are supplemented by the short-range
two-pion-exchange-contact ($2\pi$-contact) term and by the leading
relativistic corrections to 3NF \cite{3nf_n3lo_short}.
 The 3NF at N$^3$LO order
does not involve any new unknown low-energy constants (LECs) and depends only
on two parameters, $c_D$ and $c_E$ that parameterize the leading
one-pion-contact term and the 3N contact term present already at N$^2$LO.
The $c_D$ and $c_E$ values need to be then fixed at this order,
as at N$^2$LO, from a fit to few-nucleon data.
At the higher order, N$^4$LO, in addition to long- and
intermediate-range interactions generated by pion-exchange
diagrams \cite{krebs2012,krebs2013}, the chiral N$^4$LO 3NF involves thirteen 
purely short-range
operators, which have been worked out in \cite{girlanda2011}.

Since the advent of numerically exact three-nucleon continuum Faddeev 
calculations the elastic nucleon-deuteron (Nd)
 scattering and the deuteron breakup reaction  
 have become a powerful tool to test  modern models of
 the nuclear forces \cite{glo96,pisa,hanover}. 
 With the appearance of high precision (semi)phenomenological NN
 potentials and first models of 3NF the question
 about the importance of 3NF has developed into the main topic of 
 3N system  studies. That issue has been given a new impetus 
 by ChPT-based approaches  
 and the possibility to apply consistent
 two- and many-body nuclear forces, derived within this framework,
 in 3N continuum calculations.

First applications of (semi)phenomenological  NN and 3N forces 
 to elastic Nd scattering and to the nucleon-induced deuteron 
 breakup reaction revealed interesting cases of discrepancies between
 pure two-nucleon (2N) theory  and data, indicating possibly large 3NF
 effects \cite{wit2001,kuros2002}.

 Using chiral 3NF's in 3N continuum 
 requires numerous time consuming computations  with varying strengths
 of the contact terms in order to establish their values.  
 They can be determined for example
 from the $^3$H binding energy and the minimum 
 of the elastic Nd scattering  
 differential cross section at the energy ($E_{lab}\approx 70$~MeV),
 where the effects of 3NF start to emerge in elastic Nd scattering
 \cite{wit2001,wit98}.  
Specifically at N$^2$LO, after establishing the so-called ($c_D,c_E$)
correlation line,
which for a particular chiral NN potential combined
with a N$^2$LO 3NF gives pairs of ($c_D,c_E$) values reproducing 
the $^3$H binding energy, 
a fit to  experimental data for the elastic Nd cross section is performed to
determine the $c_D$ and $c_E$ strengths.
Fine-tuning of the 3N Hamiltonian parameters requires an extensive
analysis of available 3N elastic Nd scattering and breakup data. That
ambitious goal calls for a significant reduction of computer time necessary to 
solve the 3N Faddeev equations.  Thus finding an efficient emulator for exact
solutions of the 3N Faddeev equations seems to be essential and of
high priority.

In this paper we propose such an emulator and test its efficiency 
as well as ability to accurately reproduce exact solutions of 3N Faddeev
equations.  
In Sec.~\ref{section2} we briefly describe the formalism of 
3N continuum Faddeev 
calculations and present the new scheme which is well-suited to 
fast calculations with varying strengths of the contact terms
in a chiral 3NF.
Tests and an evaluation of that new approach based on example
 observables in elastic Nd 
 scattering and in selected breakup configurations
 are presented in Sec.~ \ref{res}. 
 We summarize and conclude in Sec. ~\ref{summary}.

\section{Perturbative treatment of contact terms in 3N Faddeev equations}
\label{section2}

Theoretical predictions  shown in the present paper were obtained
within the  3N Faddeev formalism using chiral two- and three-body
forces. The formalism itself and numerical performance were presented
in numerous publications so we only briefly summarize the formalism
and for details refer to ~\cite{glo96,wit88,hub97,book}.

Neutron-deuteron scattering with nucleons interacting
via NN interactions $v_{NN}$ and a 3NF $V_{123}=V^{(1)}+V^{(2)}+V^{(3)}$, is
described in terms of a breakup operator $T$ satisfying the
Faddeev-type integral equation~\cite{glo96,wit88,hub97}
\begin{eqnarray}
T\vert \phi \rangle  &=& t P \vert \phi \rangle +
(1+tG_0)V^{(1)}(1+P)\vert \phi \rangle + t P G_0 T \vert \phi \rangle \cr 
&+& 
(1+tG_0)V^{(1)}(1+P)G_0T \vert \phi \rangle \, .
\label{eq1a}
\end{eqnarray}
The 2N $t$-matrix $t$ is the solution of the
Lippmann-Schwinger equation with the interaction
$v_{NN}$.   $V^{(1)}$ is the part of a 3NF which is 
symmetric under the interchange of nucleons $2$ and $3$: $V_{123}=V^{(1)}(1+P)$.
 The permutation operator $P=P_{12}P_{23} +
P_{13}P_{23}$ is given in terms of the transposition operators,
$P_{ij}$, which interchange nucleons $i$ and $j$.  The initial state 
$\vert \phi \rangle = \vert \mathbf{q}_0 \rangle \vert \phi_d \rangle$
describes the free motion of the neutron and the deuteron 
  with the relative momentum
  $\mathbf{q}_0$  and contains the internal deuteron wave function
  $\vert \phi_d \rangle$.
Finally, $G_0$ is the resolvent of the three-body center-of-mass kinetic
energy. 
The amplitude for elastic scattering leading to the 
two-body final state $\vert \phi ' \rangle$ is then given by~\cite{glo96,hub97}
\begin{eqnarray}
\langle \phi' \vert U \vert \phi \rangle &=& \langle \phi' 
\vert PG_0^{-1} \vert 
\phi \rangle + 
\langle \phi' \vert PT \vert \phi \rangle + \langle 
\phi'\vert  V^{(1)}(1+P)\vert \phi \rangle  \cr
&+& \langle \phi' \vert V^{(1)}(1+P)G_0T\vert  \phi \rangle,
\label{eq3}
\end{eqnarray}
while the corresponding amplitude for the breakup reaction reads
\begin{eqnarray}
\langle  \vec p \vec q \vert U_0 \vert \phi \rangle &=&\langle 
 \vec p \vec q \vert  (1 + P)T\vert
 \phi \rangle ,
\label{eq3_br}
\end{eqnarray}
where the free three-body breakup channel state $\vert  \vec p \vec q \rangle $
is defined in terms of the two Jacobi (relative) momentum vectors $\vec p$
and $\vec q$. 

We solve Eq.~(\ref{eq1a}) in the momentum-space partial wave basis
$\vert p q \alpha \rangle$, determined by 
the magnitudes of the 
3N Jacobi momenta $p$ and $q$ together with the angular momenta and isospin 
quantum numbers $\alpha$ containing the 2N subsystem spin,
orbital and total angular momenta $s, l$ and $j$, the spectator nucleon orbital
and total angular momenta with respect to the center of mass (c.m.) of the 2N
subsystem, $\lambda$ and $I$:
\begin{eqnarray}
\vert p q \alpha \rangle \equiv \vert p q (ls)j (\lambda \frac {1} {2})I (jI)J
  (t \frac {1} {2})T \rangle ~.
\label{eq4a}
\end{eqnarray}
The total 2N and spectator angular momenta $j$ and $I$ as well as isospins
$t$ and $\frac {1} {2}$, are finally 
coupled to the total angular momentum $J$ and isospin $T$ of the 3N system.
In practice a converged solution of Eq.~(\ref{eq1a})
using partial wave decomposition
in momentum space at a given energy $E$ requires taking all 3N partial wave
states up to the 2N angular momentum $j_{max}=5$ 
and the 3NF force acting up to the 3N total
angular momentum $J=7/2$. The number of resulting
partial waves (equal to the number of coupled integral equations in
two continuous
variables $p$ and $q$)
amounts to $142$. The required computer time to get one solution on a 
personal computer is about
$\approx 2$~h. In the case when such calculations have to be performed for a
big number of varying 3NF parameters, time restrictions
become prohibitive. Below we propose an approximate calculational
 scheme  which enables to reduce
 significantly the required time of calculations.  

 To be specific, let us take a chiral 3NF with a number of parameters which
 are the strengths of the contact
 terms. The 3NF at N$^2$LO has one parameter-free term (2$\pi$-exchange
 contribution) and two contact terms with strength parameters $c_D$ and $c_E$.
 At N$^3$LO there are more several parameter-free parts 
  but again only two contact terms.
  At N$^4$LO again, parameter free contributions are supplemented by
  fifteen contact
 terms with strengths: $c_D$, $c_E$,  $c_{E_1}$, ..., $c_{E_{13}}$.
 All these contact terms are restricted to 
 small 3N total angular momenta and
 to only few partial wave states for a given total 3N angular momentum 
 $J$ and parity $\pi$. For example for $J^{\pi}=7/2^{\pm}$ all
 the matrix elements $< p q \alpha \vert V^{(1)}  \vert p' q' \alpha' >$
 proportional to 
 $c_{E_1}$ and $c_{E_7}$  vanish, while the $c_D$ and $c_E$ terms are nonzero only
 for a restricted number of  $\alpha, \alpha'$ pairs  
 (mostly these containing $^1S_0$ and
   $^3S_1-^3D_1$ quantum numbers) \cite{epel2002,epel_tower}.  
   Bearing that in mind and taking into account the fact that contact 
   terms yield a small
   contribution to the 3N potential energy compared to the
   leading, parameter-free part, it is possible to apply a perturbative
 approach in order to include the contact terms.

 Let us split the $V^{(1)}$ part of 3NF into a parameter-free term $V(\theta_0)$
 and a sum of contact terms $\Delta V(\theta)$:
\begin{eqnarray}
  \langle p q \alpha \vert V^{(1)}(1+P) \vert p' q' \alpha' \rangle &=&
  \langle p q \alpha \vert V(\theta_0) (1+P) \vert p' q' \alpha' \rangle \cr
  &+& \langle p q \alpha \vert \Delta V(\theta) (1+P)
  \vert p' q' \alpha' \rangle ~,
\label{eq4}
\end{eqnarray}
with $\theta_0=(c_D=0,c_E=0,c_{E_i}=0)$ and $\theta=(c_D,c_E,c_{E_i})$ the set
of values for contact terms for which we would like to find solution of
3N Faddeev equations.

Then we divide the 3N partial wave states into two sets: $\beta$ and
the remaining one, $\alpha$. The $\beta$ set is defined by nonvanishing
matrix elements
of $\Delta V(\theta)$. From Eq.~(\ref{eq1a}) one obtains (omitting the Jacobi
momenta in notation of partial wave states):
\begin{eqnarray}
  \langle \alpha \vert T (\theta) \vert \phi \rangle  &=&
 \langle \alpha \vert t P \vert \phi \rangle +
 \langle \alpha \vert (1+tG_0) [ V(\theta_0) + \Delta V(\theta) ]
 (1+P)\vert \phi  \rangle
 + \langle \alpha \vert t P G_0 T(\theta) \vert \phi \rangle \cr
 &+&  \langle \alpha \vert (1+tG_0) [V (\theta_0) + \Delta V(\theta) ]
 (1+P)G_0T(\theta) \vert \phi \rangle \cr
  \langle \beta \vert T (\theta) \vert \phi \rangle  &=&
 \langle \beta \vert t P \vert \phi \rangle +
 \langle \beta \vert (1+tG_0) [ V(\theta_0) + \Delta V(\theta) ]
 (1+P)\vert \phi  \rangle
 + \langle \beta \vert t P G_0 T(\theta) \vert \phi \rangle \cr
 &+&  \langle \beta \vert (1+tG_0) [V (\theta_0) + \Delta V(\theta) ]
 (1+P)G_0T(\theta) \vert \phi \rangle ~.
\label{eq5}
\end{eqnarray}

Introducing $T(\theta_0)$ and $\Delta T(\theta)$ such that 
 $T(\theta)=T(\theta_0) + \Delta T(\theta)$, one gets:
\begin{eqnarray}
  \langle \alpha \vert T (\theta_0) \vert \phi \rangle
+  \langle \alpha \vert \Delta T (\theta) \vert \phi \rangle
  &=&
 \langle \alpha \vert t P \vert \phi \rangle +
 \langle \alpha \vert (1+tG_0)  V(\theta_0) (1+P)\vert \phi  \rangle \cr
 &+& \langle \alpha \vert (1+tG_0)  \Delta V(\theta) (1+P)\vert \phi  \rangle
 + \langle \alpha \vert t P G_0 T(\theta_0) \vert \phi \rangle \cr
 &+& \langle \alpha \vert t P G_0 \Delta T(\theta) \vert \phi \rangle \cr
 &+& \langle \alpha \vert (1+tG_0) V (\theta_0) (1+P)G_0T(\theta_0)
  \vert \phi \rangle  \cr
  &+& \langle \alpha \vert (1+tG_0) V (\theta_0) (1+P)G_0 \Delta T(\theta)
  \vert \phi \rangle \cr
  &+& \langle \alpha \vert (1+tG_0)  \Delta V(\theta) (1+P)G_0T(\theta_0)
  \vert \phi \rangle \cr
  &+& \langle \alpha \vert (1+tG_0) \Delta V(\theta) (1+P)G_0 \Delta T(\theta)
  \vert \phi \rangle ~.
\label{eq6}
\end{eqnarray}

Since $\Delta V(\theta)$ has nonvanishing elements only
for channels $\vert \beta \rangle$  then it follows that
\begin{eqnarray}
  \langle \alpha \vert (1+tG_0)  \Delta V(\theta) (1+P)\vert \phi  \rangle
  &=&0 \cr
\langle \alpha \vert (1+tG_0)  \Delta V(\theta) (1+P)G_0T(\theta_0)
 \vert \phi \rangle &=& 0 \cr
 \langle \alpha \vert (1+tG_0) \Delta V(\theta) (1+P)G_0 \Delta T(\theta)
 \vert \phi \rangle &=& 0 ~,
\label{eq6a}
\end{eqnarray}
and  Eq.~(\ref{eq6}) can be written as two separate equations for
 $\langle \alpha \vert T (\theta_0) \vert \phi \rangle$ and
 $\langle \alpha \vert \Delta T (\theta) \vert \phi \rangle$:
\begin{eqnarray}
  \langle \alpha \vert T (\theta_0) \vert \phi \rangle
  &=&
 \langle \alpha \vert t P \vert \phi \rangle +
 \langle \alpha \vert (1+tG_0)  V(\theta_0) (1+P)\vert \phi  \rangle 
 + \langle \alpha \vert t P G_0 T(\theta_0) \vert \phi \rangle \cr
 &+& \langle \alpha \vert (1+tG_0) V (\theta_0) (1+P)G_0T(\theta_0)
  \vert \phi \rangle  \cr
  \langle \alpha \vert \Delta T (\theta) \vert \phi \rangle &=&
   \langle \alpha \vert t P G_0 \Delta T(\theta) \vert \phi \rangle
   + \langle \alpha \vert (1+tG_0) V (\theta_0) (1+P)G_0 \Delta T(\theta)
  \vert \phi \rangle ~.
\label{eq7}
\end{eqnarray}
 
Inserting the decomposition of $T(\theta)$ into the second equation in
(\ref{eq5}) for channels
$\vert \beta \rangle$ one obtains:
\begin{eqnarray}
  \langle \beta \vert T (\theta_0) \vert \phi \rangle
  &=&
 \langle \beta \vert t P \vert \phi \rangle +
 \langle \beta \vert (1+tG_0)  V(\theta_0) (1+P)\vert \phi  \rangle 
 + \langle \beta \vert t P G_0 T(\theta_0) \vert \phi \rangle \cr
 &+& \langle \beta \vert (1+tG_0) V (\theta_0) (1+P)G_0T(\theta_0)
  \vert \phi \rangle  \cr
  \langle \beta \vert \Delta T (\theta) \vert \phi \rangle &=&
 \langle \beta \vert (1+tG_0) \Delta V(\theta) (1+P)\vert \phi  \rangle
 + \langle \beta \vert (1+tG_0) \Delta V(\theta) (1+P)G_0 T(\theta_0)
 \vert \phi  \rangle \cr
 &+& \langle \beta \vert (1+tG_0)[ V (\theta_0) + \Delta V(\theta)]
 (1+P)G_0 \Delta T(\theta)   \vert \phi \rangle  \cr
 &+&  \langle \beta \vert t P G_0 \Delta T(\theta) \vert \phi \rangle ~.
\label{eq8}
\end{eqnarray}

The first equations in (\ref{eq7}) and (\ref{eq8}) are the
Faddeev equations (\ref{eq1a}) for $T(\theta_0)$. Since the two leading terms
for $\langle \beta \vert \Delta T (\theta) \vert \phi \rangle$ in (\ref{eq8})
 are of the order
 of $\Delta V(\theta)$ then
$\langle \alpha \vert \Delta T (\theta) \vert \phi \rangle \approx 0$ and
the calculations proceed so that in the first step a solution for
$T(\theta_0)$ is found (it is independent from parameters $\theta$).
In the next step a solution of second equation in the set (\ref{eq8}) for
$\langle \beta \vert \Delta T (\theta) \vert \phi \rangle$ is obtained and from
that $\langle \alpha \vert \Delta T (\theta) \vert \phi \rangle$ is
calculated  by:
\begin{eqnarray}
\langle \alpha \vert \Delta T (\theta) \vert \phi \rangle &=&
\langle \alpha \vert t P G_0  \sum_{\beta} \int_{p'q'}
\vert p' q' \beta \rangle \langle p' q' \beta \vert
\Delta T(\theta)
 \vert \phi \rangle \cr 
&+& \langle \alpha \vert (1+tG_0) V (\theta_0) (1+P)G_0
\sum_{\beta} \int_{p'q'}  \vert p'q' \beta \rangle
\langle p' q' \beta \vert   \Delta T(\theta)
  \vert \phi \rangle ~.
\label{eq9}
\end{eqnarray}

Finally, $T(\theta)$ is calculated as
\begin{eqnarray}
  \langle \alpha \vert T (\theta) \vert \phi \rangle &=&
  \langle \alpha \vert T (\theta_0) \vert \phi \rangle
  +  \langle \alpha \vert \Delta T (\theta) \vert \phi \rangle  \cr
  \langle \beta \vert T (\theta) \vert \phi \rangle &=&
  \langle \beta \vert T (\theta_0) \vert \phi \rangle +
  \langle \beta \vert \Delta T (\theta) \vert \phi \rangle
\label{eq10}
\end{eqnarray}

\section{Results}
\label{res}

To check the quality of the proposed scheme we have chosen
one NN potential from among available
chiral NN interactions, namely  
the SMS N$^4$LO$^+$ chiral potential of the
Bochum-Bonn group \cite{preinert} with  the regularization
 cutoff $\Lambda = 450$~MeV, and combined it with the chiral N$^2$LO
 3NF augmented by one out of thirteen contact terms from the N$^4$LO 3NF,
 that is $E_7$.  
 The low-energy constants  
of the contact interactions in that 3NF were adjusted to the 
 triton binding energy 
 and we used  the set of strengths
  $\theta=(c_D=-8.2053,c_E=-1.0019,c_{E_7}=2.0)$
 (according to the notation of Refs.~\cite{epel2002,epel_tower}).

 We solved the 3N Faddeev equation (\ref{eq1a}) exactly at two incoming
 neutron energies 
$E=70$ and $190$~MeV with such a choice of
NN and 3N forces
as well as with the same NN potential combined with the 3NF restricted only to
the
parameter free $2\pi$-exchange N$^2$LO term  
(set $\theta_0=(c_D=0.0,c_E=0.0,c_{E_7}=0.0)$).
The first energy was taken from a region
where 3NF effects start to appear in 3N continuum observables
and the second one from a range with well-developed 3NF
effects \cite{wit2001,kuros2002,wit98}. We emphasize that the
solution with the $\theta_0$ set forms a starting point in the proposed
perturbative treatment  of Eqs.~(\ref{eq7})-(\ref{eq10}) and has to be
calculated only once, regardless of how many variations of strength parameters
are required. 
In the next step, we performed, at the same energies,  
 our perturbative  treatment,  solving first the second
equation in set (\ref{eq8}). 
Having determined $\langle \alpha \vert \Delta T (\theta) \vert \phi \rangle$
from  Eq.(\ref{eq9}) we calculated our emulator solution of Eq.~(\ref{eq10}), 
 with two sets of 3N channels $ \vert \beta \rangle$, one
 comprising the 2N subsystem states $^1S_0+^3S_1-^3D_1$ and second
 including all 2N states with the total angular momentum $ j\le 2$. 
 The number of 3N partial waves diminishes from $142$ to $34$ in the second 
 set of states and to $20$ in the first one.
 Such a smaller set of 3N channels
 leads to a reduction (by a factor of $\approx 4$) of the 
 computer time required in
 the perturbative approach as compared to the exact computation 
 (one run in the exact approach requires $\approx 30$~minutes of a personal 
 computer time, provided that the 
 $V(\theta_0) (1+P)$ and $V(\theta_i) (1+P)$ 
 kernels, acting in 
  $ (1+tG_0) V (\theta) (1+P)G_0T(\theta)
  \vert \phi \rangle$ term of Eq.~(\ref{eq1a}), 
 are calculated in advance  (with the strength of the
 contact term $i: c_i$ and   $\theta_i=(c_i=1.0,c_{k\ne i}=0.0) $). 

\begin{figure}    
\includegraphics[scale=0.62,clip=true]{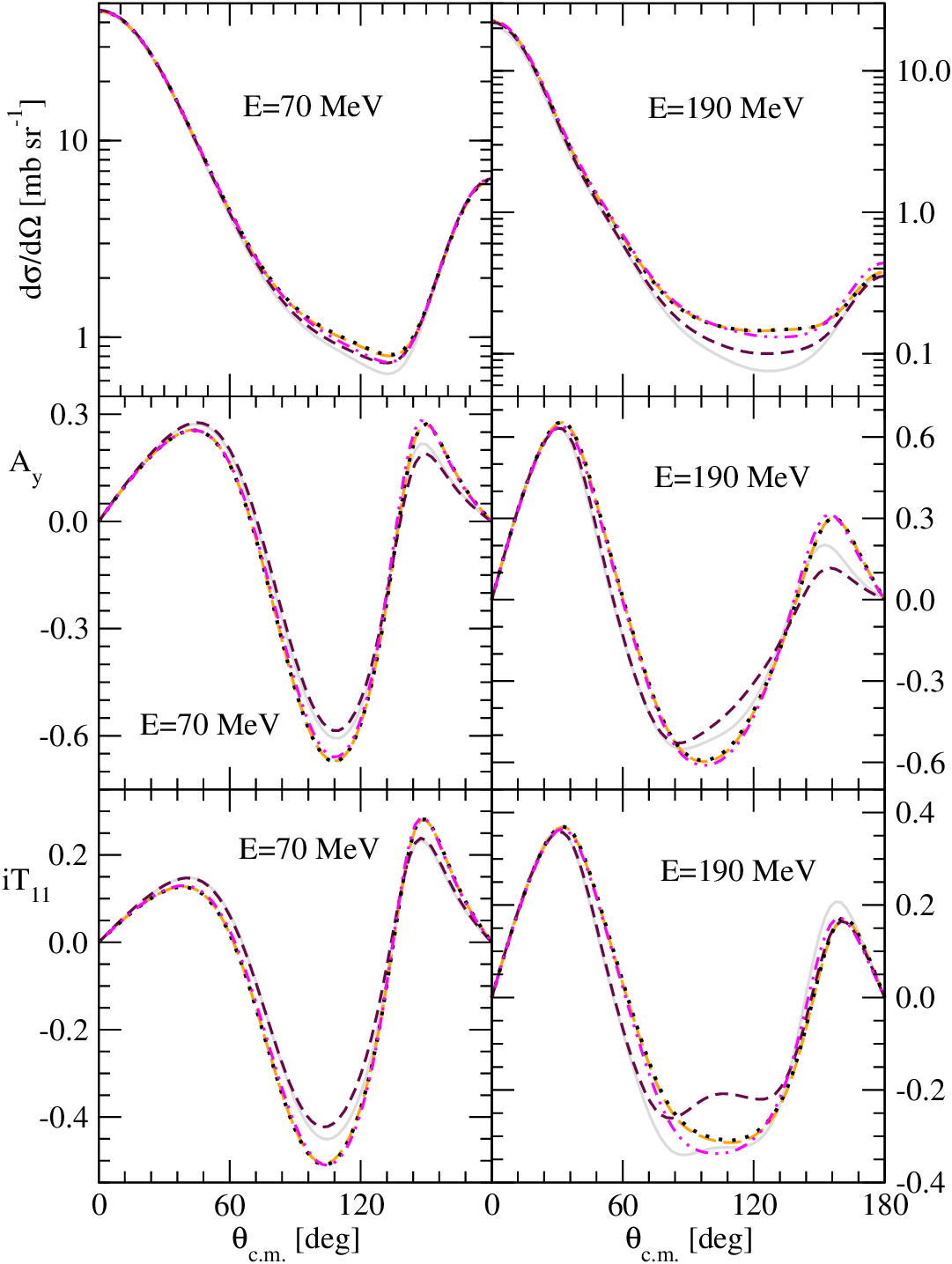}
\caption{
(color online) 
  The angular distribution $\frac {d\sigma} {d\Omega}$,
  the neutron analyzing power $A_y$, and the 
  deuteron vector analyzing power $iT_{11}$ in elastic nd scattering
  at the incoming
  neutron laboratory energy $E=70$ and $190$~MeV.
  The different curves are predictions of the chiral SMS N$^4$LO$^+$ NN
  potential with
  regulator $\Lambda=450$~MeV alone ((grey) solid line) or combined with the
  N$^2$LO 3NF comprising also  one contact term ${E_7}$ from
  N$^4$LO. The exact prediction for that combination with strength parameters
  ($c_D=-8.2053,c_E=-1.0019,c_{E_7}=2.0$) is shown by the (orange)
  long dashed line. The exact predictions for parameter free term in N$^2$LO
  3NF (set of parameters ($c_D=0.0,c_E=0.0,c_{E_7}=0.0$) is
  shown by (maroon) short dashed line. 
  The predictions based on approximate
  treatment with channels $\vert \beta \rangle$ restricted
  to only $^1S_0+^3S_1-^3D_1$ and to all $j_{max}=2$ channels,  
  are represented by (magenta) dash-double-dotted and (black) dotted lines,
  respectively.
}
\label{fig1}
\end{figure}

\begin{figure}    
\includegraphics[scale=0.62,clip=true]{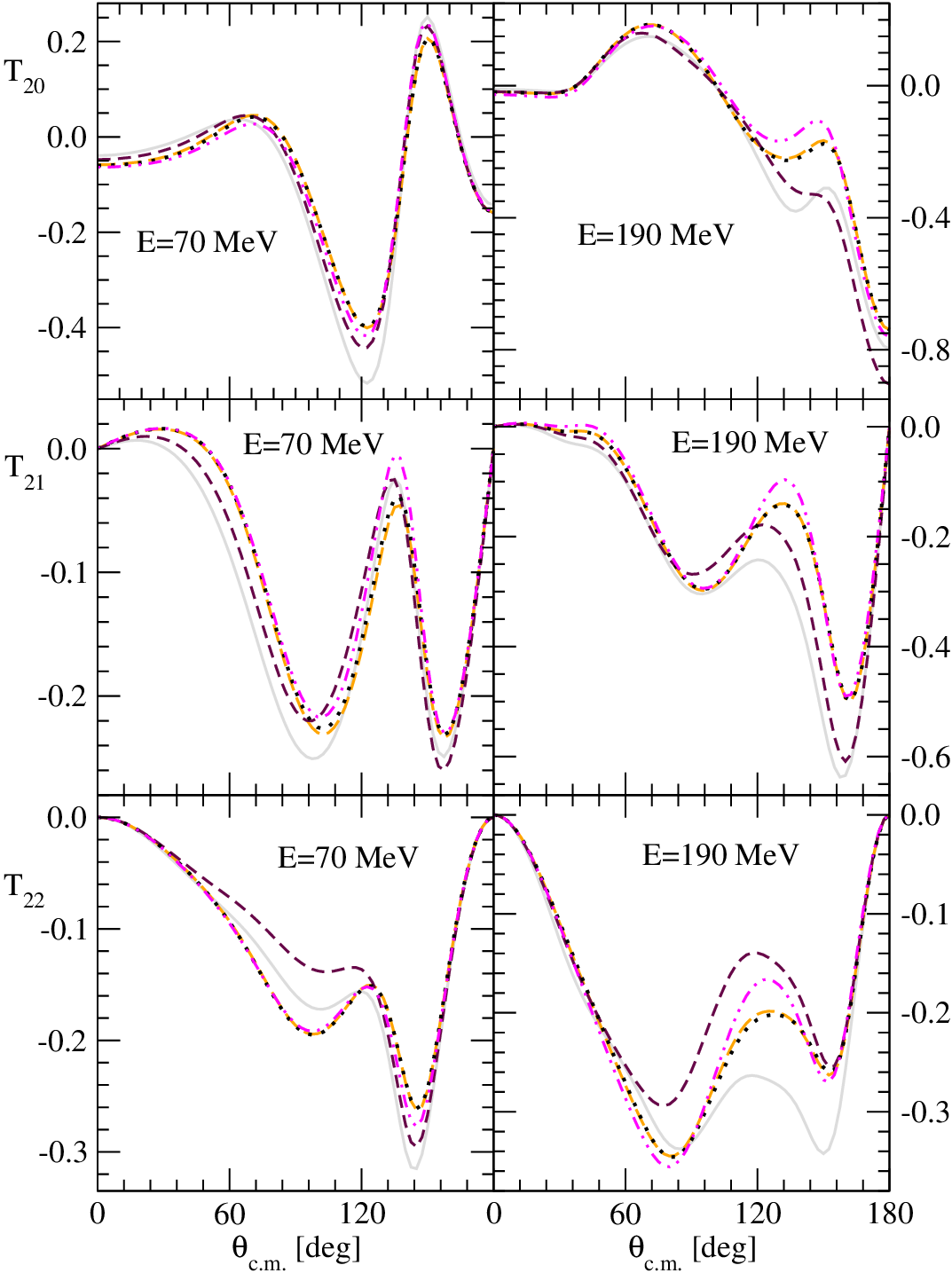}
\caption{
(color online) 
  The same as in Fig.~\ref{fig1} but for deuteron tensor analyzing powers
  $T_{20}, T_{21}$, and $T_{22}$.
}
\label{fig2}
\end{figure}

\begin{figure}   
\includegraphics[scale=0.62,clip=true]{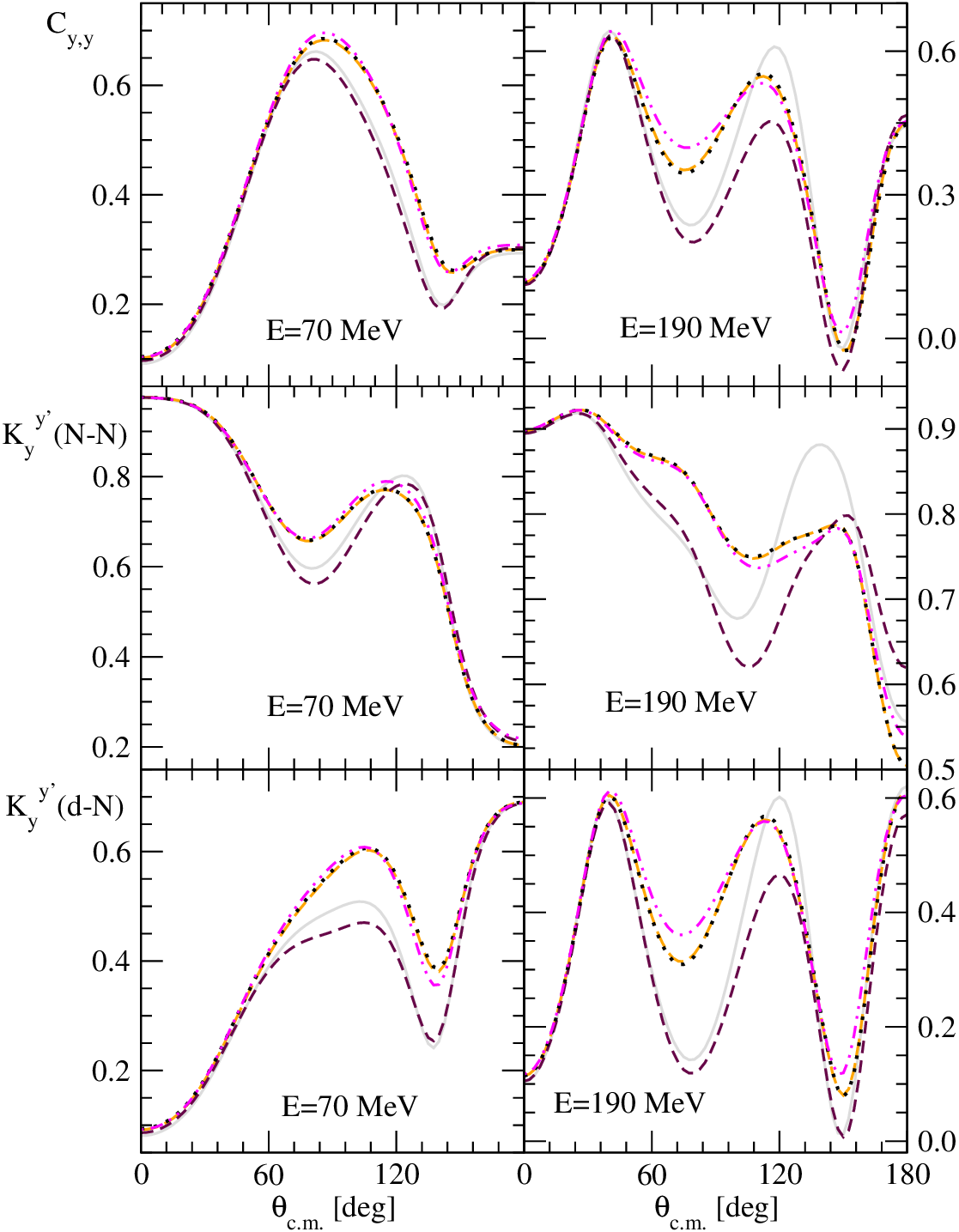}
\caption{
(color online) 
  The same as in Fig.~\ref{fig1} but for the spin correlation $C_{y,y}$, 
  polarization  transfer coefficient from nucleon to nucleon $K_y^{y'}(N-N)$,
  and polarization  transfer coefficient from deuteron to nucleon 
  $K_y^{y'}(d-N)$.
}
\label{fig3}
\end{figure}

 In Figs.~\ref{fig1}-\ref{fig3} we compare results of the perturbative treatment
 with the exact one for a number of neutron-deuteron (nd) elastic
 scattering observables (the differential cross
 section, the nucleon and deuteron vector analyzing powers in Fig.~\ref{fig1},
 the deuteron tensor analyzing powers in Fig.~\ref{fig2}, and the
  spin correlation
 and polarization transfer coefficients in Fig.~\ref{fig3}).
 The perturbative approach reproduces the exact predictions with
 $\approx 1$~\% accuracy.
 Even the restricted to $^1S_0+^3S_1-^3D_1$ choice 
 of $\vert \beta \rangle$ channels ((magenta) dash-double-dotted line)
   follows quite well the exact predictions ((orange) long dashed line). 
   The results for the choice $j \le 2$ ((black) dotted line) are practically
   indistinguishable from the exact ones. We checked that this picture
   remains the
   same for all other (not shown) elastic scattering observables
   (altogether $55$, comprising in addition to the presented ones,
   also all the 
   remaining spin correlation- and polarization transfer-coefficients
   between all participating particles).

\begin{figure}  
\includegraphics[scale=0.65,clip=true]{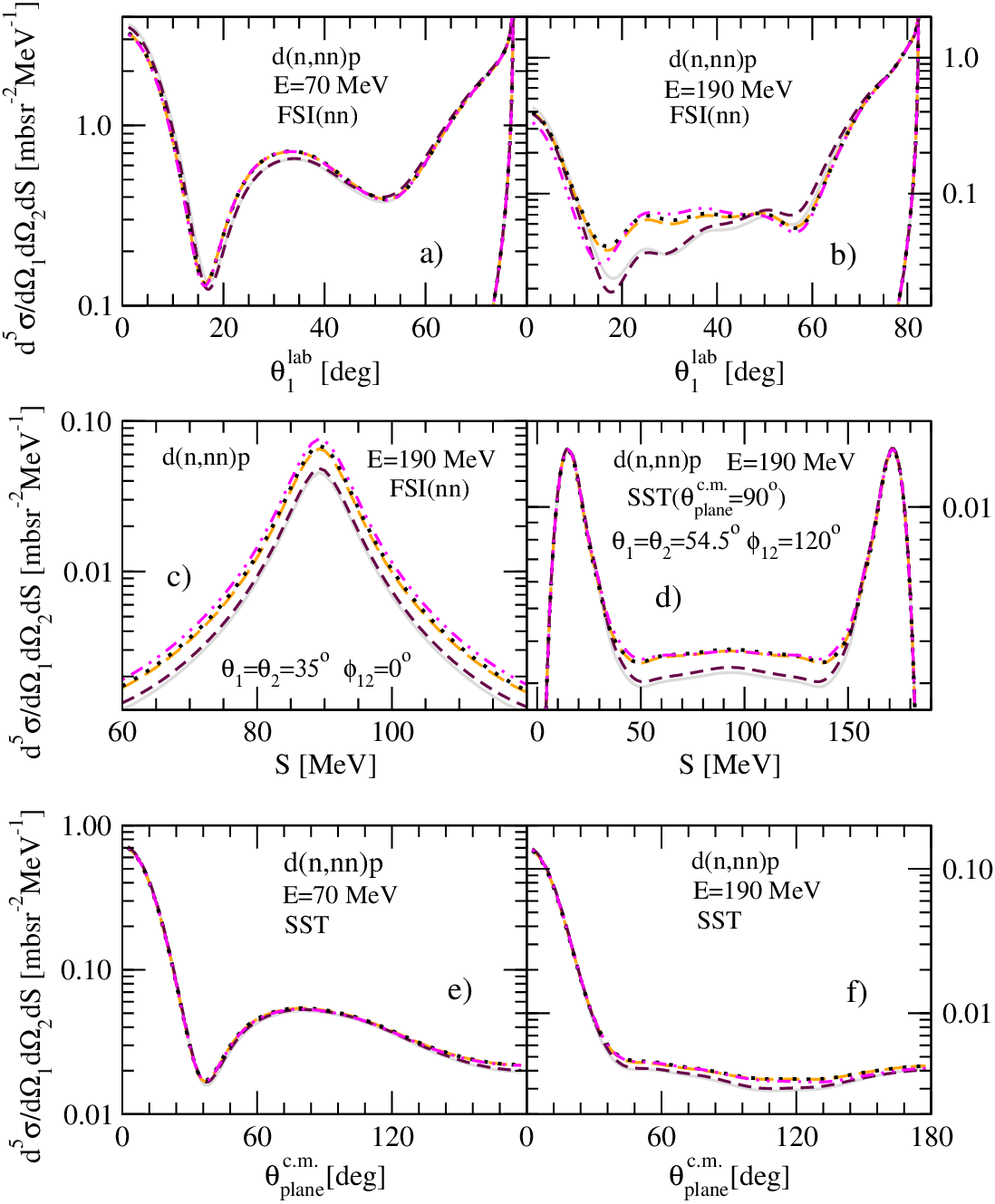}
\caption{
(color online) 
The exclusive breakup d(n,nn)p 
cross section $d^5\sigma/d\Omega_1d\Omega_2dS$ 
 at the incoming neutron lab. energy $E=70$ and $190$~MeV 
 for the neutron-neutron FSI  condition  ($\vec p_1=\vec p_2$)
  as a function of the lab. 
 production angle of the outgoing 
 neutrons $\theta_1^{lab}=\theta_2^{lab}$ and $\phi_{12}=0^o$: a), b), and for 
 the SST condition  as a function of the c.m. angle 
 between   plane,  in which in the 3N c.m. frame momenta of three outgoing
 nucleons are placed,  and  the incoming nucleon momentum: e), f). In c) 
 the cross sections for a particular FSI(nn) configuration from b)  
 produced at  $\theta_1^{lab}=\theta_2^{lab}=35^o$ and in
 d) for the SST from f) with $\theta_{plane}^{~c.m.}=90^o$ 
 are  shown along the arc-length S.  
 For the description of lines see Fig.~\ref{fig1}.
}
\label{fig4}
\end{figure}

   In Fig.~\ref{fig4} we show the results for breakup cross sections in
   selected kinematically
   complete breakup configurations, namely for the final-state-interaction (FSI)
   and symmetrical-space-star (SST) geometries. 
 In the FSI configuration under the exact FSI condition the two outgoing 
 nucleons have equal momenta and strongly interact in
 the $^1S_0$ state. 
 We show in Fig.~\ref{fig4}a and b the cross section
for FSI(1-2) in the d(n,nn)p breakup reaction 
exactly at the FSI condition as a function of 
the  laboratory angle of one of the two FSI interacting nucleons.
 For one detection angle we show in Fig.~\ref{fig4}c  
 also the FSI cross section along the arc-length parameter S, 
 which for the fixed angles defines unambiguously the energies of all the three
 outgoing nucleons.

 Under the symmetrical-space-star condition
 the momenta of the free outgoing nucleons have the same magnitudes 
 in the 3N c.m. frame
 and form a three-pointed "Mercedes-Benz" star in the plane symmetrical
 with respect to the incoming nucleon momentum
 (the plane is bent at the angle $\theta_{plane}^{~c.m.}$ 
 with respect to the incoming nucleon momentum). 
 We show the SST cross section 
 as a function of that angle in Fig.~\ref{fig4}e and f as well as for
 the case of $\theta_{plane}^{~c.m.}=90^o$
along the arc-length parameter S in Fig.~\ref{fig4}d.
 The results are similar to the elastic scattering case. Again the perturbative
treatment reproduces quite well cross sections calculated with exact
solutions. Only at $190$~MeV in the region of FSI peak the precision is reduced 
  and amounts to $\approx 2-4$~\%.

\section{Summary and conclusions}
\label{summary}

We presented an approximate approach  which enables us to take 
into account contact terms of a chiral 3NF in the 3N continuum Faddeev
calculations. 
Such contact terms are short ranged and thus act only in the partial
waves with low angular momenta. That  reduction of 3N
partial wave states together with small magnitudes of the contact contributions
 as compared with 
the leading NN potential and parameter-free terms in a 3NF  enable us 
 to treat the contact terms perturbatively. 
 The proposed perturbative  approach allows one to reduce
 by the factor of about four the required computation time and is thus
 especially suited to repeated calculations with varying strengths of
  contact terms. 
 We checked that the proposed treatment allows us to reproduce surprisingly well
the exact predictions for nd elastic scattering as well as for
nd breakup observables. 
It is conceivable that with the help of the constructed emulator of
 the exact solutions of 3N Faddeev equations  fine
 tuning of a 3N Hamiltonian parameters based on available 3N scattering data
 is feasible.

\begin{acknowledgements}
This study has been performed within Low Energy Nuclear Physics
International Collaboration (LENPIC) project and 
was  supported by the Polish National Science Center 
 under Grant No. 2016/22/M/ST2/00173. 
 The numerical calculations were performed on the 
 supercomputer cluster of the JSC, J\"ulich, Germany.
 We would like to thank A. Nogga, K. Hebeler, and P. Reinert for providing us
 with matrix elements of chiral 3NF's.
\end{acknowledgements}

%
%
%
%
%
\vfill.
\newpage

\end{document}